\documentclass[12pt,preprint]{aastex}

\shorttitle{METALLICITY AND AGN ACTIVITY ON MIR EMISSION}
\shortauthors{Wu, et al.}

\begin{document}

\title{EFFECTS OF METALLICITY AND AGN ACTIVITY ON THE MID-INFRARED DUST
EMISSION OF GALAXIES}

\author{
Hong Wu\altaffilmark{1,2},
Yi-Nan Zhu\altaffilmark{1,3},
Chen Cao\altaffilmark{1,3},
Bo Qin\altaffilmark{1}
}

\email{hwu@bao.ac.cn}
\altaffiltext{1}{National Astronomical Observatories, Chinese Academy of Sciences, A20 Datun Road, Beijing 100012, P.R.\ China}
\altaffiltext{2}{Visiting Scholar, Institute for Astronomy, University of Hawaii, 2680 Woodlawn Drive, Honolulu, HI 96822}
\altaffiltext{3}{Graduate School, Chinese Academy of Sciences, Beijing, 100039, P.R.\ China}

\begin{abstract}
Using a sample of the $Spitzer$ SWIRE-field galaxies whose optical
spectra are taken from the Data Release 4 of the Sloan Digital Sky Survey, 
we study possible correlations between the Mid-InfraRed (MIR) dust emission from these
galaxies and both their metallicities and AGN activities. We find that
both metallicity and AGN activity are well correlated with the
following ratios: PAH(Polycyclic Aromatic Hydrocarbon)-to-star, 
VSG(Very Small Grain)-to-star and PAH-to-VSG, which can
be characterized by $\nu L_{\nu}[8\mu m({\rm dust})]/\nu L_{\nu}[3.6\mu m]$, 
$\nu L_{\nu}[24\mu m]/\nu L_{\nu}[3.6\mu m]$, and
$\nu L_{\nu}[8\mu m({\rm dust})]/\nu L_{\nu}[24\mu m]$ respectively.
We argue that our MIR-metallicity correlation could be explained by
either the amount of dust (ongoing dust formation) or dust destruction 
(PAHs and VSGs could be destroyed by hard and intense radiation fields), 
and that the MIR-AGN correlation could arise due to either PAH destruction 
or an enhanced VSG continuum by the central AGN.

\end{abstract}


\section{INTRODUCTION}
\label{sec intro}

In galaxies, the MIR bands not only contain a large number of
atomic, ionic and molecular lines, but also cover various dust
features from grains of different size \citep{sturm02}. These
include the features related to the polycyclic aromatic hydrocarbons (PAHs) 
and the dust continuum from the very small grains (VSGs).
Recent work
\citep{calzetti05,wu05,vogler05,alonso06,elbaz02,boselli04,forster04,roussel01}
have shown that the MIR dust emission are well correlated with the star
formation rate (SFR) indicators, such as H$\alpha$, Pa$\alpha$,
radio, UV, and far-infrared in either star forming regions in
galaxies or galaxies themselves as a whole. With further advantages
of having less extinction, the MIR dust luminosities are ideal
tracers of star formation and can be widely used to estimate the SFR.

The MIR emission in galaxies could, however, be affected by many
factors. For example, early observations on the ground as well as
recent space observations by the Infrared Space Observatory ($ISO$)
and the $Spitzer$ Space Telescope \citep{werner04} have shown that the PAH
emission is generally suppressed in low-metallicity galaxies
\citep{roche91,madden00,galliano05,madden06,wuyl06}, or even
missing in the most metal-poor galaxies
\citep{houck04b,engelbracht05,thuan1999}. \citet{siebenmorgen04} have
demonstrated the absence of PAHs in the nuclei of AGN-hosting
galaxies. Using the $Spitzer$ Infrared Spectrograph \citep[IRS,][]{houck04a} spectra, 
 \citet{weedman05} have confirmed that the PAH
emission is much weaker in AGNs.
Moreover, the AGN activity could also steepen the shape of the MIR
VSG continuum in some active galaxies \citep{verma2005}.

To quantitatively examine the factors that could affect the MIR
emission in galaxies, we need a sample of galaxies with both MIR
measurements and optical spectra. The {\it Spitzer} Wide-area
InfraRed Extragalactic survey \citep[SWIRE;][]{lonsdale03} and the
Sloan Digital Sky Survey \citep[SDSS;][]{stoughton02} provide 
such opportunities. The Infrared Array Camera \citep[IRAC,][]{fazio04} 8$\mu$m band, which
just covers the main PAH features, can be used to trace the PAH
emission, while the Multiband Imaging Photometer \citep[MIPS,][]{rieke04} 24$\mu$m band, which
covers the continuum, can trace the VSG emission. The SDSS spectra
provide us information of both the metallicities and AGN activities
of galaxies. In the present paper, we mainly focus on the
correlations between the MIR emission and metallicity/AGN-activity.

In Section 2, we describe the data, data reduction, and the sample.
The spectral classification, the metallicity and the definition of
AGN activity are presented in Section 3. The correlation
analysis between the MIR and metallicity/AGN-activity are shown in
Section 4.  Finaly, Section 5 and 6 are discussions and summary.
Throughout the paper we adopt a $\Lambda$CDM cosmology
with $\Omega_M=0.3$, $\Omega_{\Lambda}=0.7$, and $H_0=70$.

\section{DATA AND SAMPLE}
The $Spitzer$ SWIRE fields \citep{lonsdale03} of ELAIS-N1, ELAIS-N2,
and Lockman Hole have been imaged by both instruments: IRAC
\citep[3.6, 4.5, 5.8, 8.0~$\mu$m;][]{fazio04} and MIPS
\citep[24$\mu$m;][]{rieke04}. These fields have also been completely
or partly covered  by the SDSS Data Release 4
\citep[DR4;][]{adelman06}. The overlap area between the three SWIRE
fields and the SDSS DR4 field used in this paper is about 15 degree$^2$.
The Basic Calibrated Data (BCD) images of the IRAC four bands were
obtained from the {\it Spitzer} Science Center, which include
flat-field corrections, dark subtraction, linearity and flux
calibrations \citep{fazio04}. The IRAC images (in all four IRAC
bands) were mosaiced from the BCD images after pointing refinement,
distortion correction and cosmic-ray removal with a final pixel
scale of 0.6\arcsec \ as described by \citet{huang04} and
\citet{wu05}, whilst the MIPS 24$\mu$m images were in a similar way
but with a final pixel scale of 1.225\arcsec \citep{wen07,cao07}. The
FWHMs of point spread functions (PSFs) in mosaiced images are around
2\arcsec \ for the IRAC four bands and 6\arcsec \ for the MIPS
24$\mu$m band. Matching the sources detected by the SExtractor
\citep{bertin96} in all the five bands with the 2MASS sources
 \citep{cutri03} ensured the astrometric uncertainties to be less than 0.1$\arcsec$.

The MIR photometries were obtained by the SExtractor with an
aperture of 6\arcsec \ for the IRAC four bands and the MIPS 24$\mu$m
band. The apertures of 6\arcsec \ were used to match the FWHM of the PSF
of the MIPS 24$\mu$m band, though the fiber aperture of the SDSS spectra is
3\arcsec. All these magnitudes are in the AB magnitude system
\citep{oke83}. The values of  -0.20, -0.17, -0.23, -0.41 magnitudes
for 3.6, 4.5, 5.8, 8.0~$\mu$m bands were adopted to correct the
apertures of 6\arcsec \ to 24\arcsec \ respectively. Comparing with
the model colors of the IRAC magnitudes and 2MASS $K_S$ magnitude
\citep{cutri03} for the stars with $J$-$K_S$ $\le$ 0.3 as
\citet{lacy05} did for sources in the {\it Spitzer} extragalactic First
Look Survey (xFLS) field, small additional corrections were made
for all the four IRAC bands. And these give us the calibration
errors in the four IRAC bands better than 0.08 mag. A factor of
-1.21 magnitude was adopted to correct the aperture from 6\arcsec \
to 30\arcsec \ for the 24$\mu$m band. An additional correction of -0.196
mag was also performed for the 24$\mu$m sources. This included a
factor of 1.15 of calibration and a factor of 1.04 of color
correction \citep{surace05}. The calibration accuracy of the 24$\mu$m photometry 
is better than 10\% \citep{rieke04}.

To build our galaxy sample, we first selected the galaxies of three
SWIRE fields from the main galaxy sample of the SDSS-DR4
\citep{adelman06} and then matched them with the MIR sources
detected in the $Spitzer$ SWIRE fields with a matching radius of
2$\arcsec$. 14 out of 1628 (about 1\%) of the SDSS galaxies were
mis-matched with the MIR sources, due to the 2\arcsec \ matching
radius. To obtain reliable line ratios and to avoid possible
contaminations from the old stellar population, we only kept the sources
with a reliable H$\beta$ flux (S/N $\ge 5$) and a higher H$
{\alpha}$ equivalent width (less than $-5.0$, the minus here means
emission), where the H$ {\alpha}$ line emission flux used here was
from BC03 \citep{bc03} continuum-subtracted. \footnote{MPA emission
line catalog, which is available at {\it
http://www.mpa-garching.mpg.de/SDSS/DR4/raw\ data.html}} This gives
our sample of 600 galaxies. All these galaxies have 8$\mu$m fluxes,
and most have 24$\mu$m fluxes. The redshifts of the galaxies in
the sample are from 0.004 to 0.18.

\section{SPECTRAL CLASSIFICATION, METALLICITY AND AGN ACTIVITY}

\begin{figure}
\centerline{\includegraphics[height=0.9\textwidth]{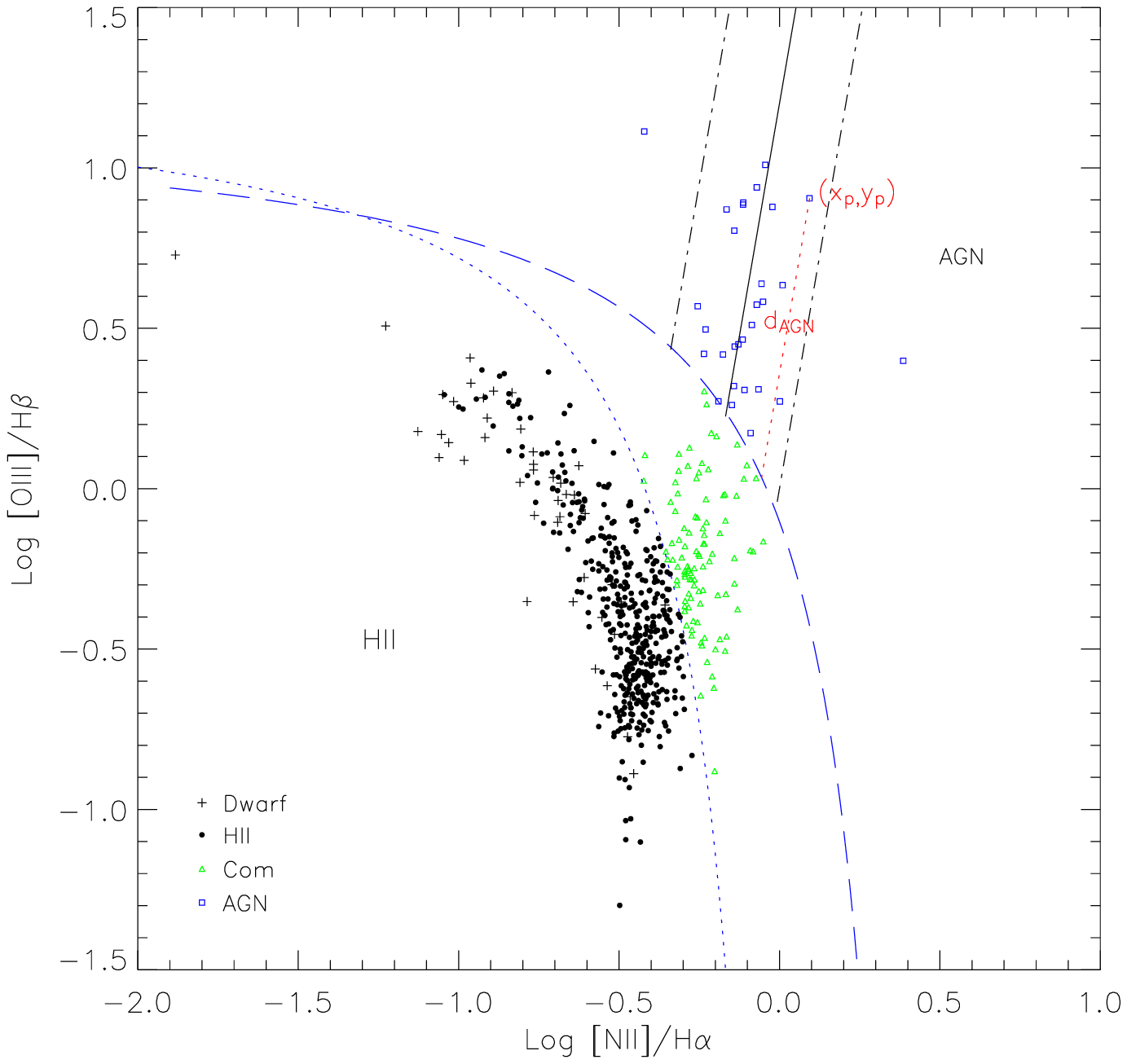}}
\caption{The line-diagnostic diagram [OIII]/H$\beta$ versus [NII]/H$\alpha$. 
The dotted curve is the parametrized curve by
\citet{kauffmann03} and the dashed curve is the theoretical boundary
for starburst \citep{kewley01}. Under the dotted curve, the HII galaxies 
are labelled as solid circles and dwarf galaxies as pluses.
The composites are between the dotted and dashed curves and
are labelled as open triangles. The AGNs are above the dashed curve
and are labelled as open boxes. To characterize AGN activity, we
define the AGN distance $d_{AGN}$ as the distance from these boxes
to the dashed curve along the best-fit solid line.} \label{fig1}
\end{figure}

Our spectral classification was based on the traditional
line-diagnostic diagram [OIII]/H$\beta$ versus [NII]/H$\alpha$,
\citep{veilleux87} as in Figure~\ref{fig1}. The dotted curve in the
figure is the parametrized curve by \citet{kauffmann03} while the
dashed curve is the theoretical boundary for starburst by
\citet{kewley01}. The galaxies below the dotted curve are
HII galaxies, the galaxies between the dotted curve and dashed curve are 
mixture types \citep{wu98} or composites \citep{kewley06}, and
galaxies above the dashed curve are narrow line AGNs. There are
altogether 472 HII galaxies, 99 composites and 29 AGNs.

The metallicities of HII galaxies were from the metallicity catalog
of SDSS DR4 \citep{tremonti04} \footnote{Available at {\it
http://www.mpa-garching.mpg.de/SDSS/DR4/Data/oh\ catalogue.html}}.
Here we adopted the median of the likelihood distribution for 12 + Log
O/H. 329 HII galaxies, which have measured metallicities in the
\citet{tremonti04} catalog, were selected for the following
analysis. 297 of them have 24$\mu$m fluxes. We defined dwarf
galaxies as the galaxies whose absolute B magnitudes were greater
than $-18$. Here the B magnitudes were calculated from the SDSS $g$-
and $r$- magnitudes \citep{smith02}. The dwarf galaxies are marked
with pluses in Figure~\ref{fig1}.

Figure~\ref{fig2} shows the metallicities of HII galaxies as a
function of the internal reddening. 
The internal reddening (characterized by E(B-V))
was derived from the Balmer decrement  H$ {\alpha}$/H$ {\beta}$
\citep{calzetti01}. The figure suggests that dusty galaxies tend to
be metal rich.

To quantify AGN activity, here we define a distance $d_{AGN}$ in dex
(see the red dotted line in Figure 1 as an example) in the traditional
line-diagnostic diagram [OIII]/H$ {\beta}$ versus [NII]/H$ {\alpha}$
to represent AGN activity. This is quite similar to the star-forming
distance defined by \citet{kewley06} in the diagnostic diagram
[OIII]/H$ {\beta}$ versus [OI]/H${\alpha}$ or [SII]/H${\alpha}$.
Considering that the composites could be dominated by star formation
and that this would weaken possible contributions from the central AGN, 
the $d_{AGN}$ is defined as the distance of an AGN from its position ($x_p$,$y_p$)
to the \citet{kewley01}'s curve along the parallel of the best linear 
fitting of all AGNs (solid line) in Figure~\ref{fig1}, and
can be formulized as :
\begin{equation}
d_{AGN} = 0.507 ( 5.8 x_p + y_p - 3.196 +A ) 
\end{equation}
\begin{equation}
A = \sqrt{14.152 + ( y_p - 5.8 x_p + 1.536 )^2},
\end{equation}
where $x=log$[NII]/H$\alpha$ and $y=log$[OIII]/H$\beta$.
Larger distances correspond to stronger AGN
activities. After removing the two most deviated AGNs, 27 AGNs between
the two dash-dotted lines were selected for our following analysis. Three
of them do not have 24$\mu$m fluxes.

\section{ANALYSIS}

\subsection{MIR Ratios Versus Metallicity}

To correct the redshift effect, we adopted the SED of the normal HII
galaxy NGC~3351 from IRS observations in the $Spitzer$ Legacy
Program SINGS \citep{kennicutt03} as the template for the
K-corrections for our sample galaxies in both the 24$\mu$m and
8$\mu$m bands \citep{wu05}. In order to avoid the strong PAH
emission, we used the 3.6$\mu$m band to estimate the stellar
contribution in the 8.0$\mu$m band. A factor of 0.26 \citep{wu05} was
used to scale the stellar continuum of 3.6$\mu$m to that of 8$\mu$m.
After subtracting the corresponding stellar continuum contribution,
we obtained the flux of 8$\mu$m dust emission denoted by
8$\mu$m(dust), as in \citet{wu05}. The 8$\mu$m(dust) includes the
dust emission from both the PAHs  and VSGs. Considering that the
contribution of the stellar continuum in the 24$\mu$m band is quite
small and negligible, we have not removed the stellar continuum from
the 24$\mu$m band flux.

It is well known that the mass of a galaxy is well related to its
metallicity \citep{pagel81}. To avoid possible effects caused by
different mass variations, we used the MIR ratios $\nu L_{\nu}[8\mu
m(dust)]/\nu L_{\nu}[3.6\mu m]$ and $\nu L_{\nu}[24\mu m]/\nu L_ {\nu}[3.6\mu m]$ 
instead of the MIR luminosities in our analysis.
These ratios can be explained as the MIR dust emission in mass unit
\citep{wen07}, since the 3.6$\mu$m luminosity can approximately
represent the stellar mass \citep{smithbj07,hancock07}.

Figure~\ref{fig3} shows the MIR ratios $\nu L_{\nu}[8\mu
m(dust)]/\nu L_{\nu}[3.6\mu m]$, $\nu L_{\nu}[24\mu m]/\nu L_{\nu}[3.6 \mu m]$, 
and $\nu L_{\nu}[8 \mu m(dust)]/\nu L_{\nu}[24
\mu m]$ as functions of metallicity in the log-log space. We plotted
the mean values (in diamonds) and the 1-$\sigma$ standard deviation
bars with a metallicity bin of 0.2. The plus symbols represent the
dwarf galaxies. We can see that the MIR ratios and metallicity are
correlated. Both $\nu L_{\nu}[8 \mu m(dust)]/\nu L_{\nu}[3.6 \mu m]$
and $\nu L_{\nu}[24 \mu m]/\nu L_{\nu}[3.6 \mu m]$ ratios show
nearly linear correlations with metallicity in the log-log space,
and the Spearman correlations are tight with the probabilities of
null-hypothesis being 2.7$\times$10$^{-17}$ and 5.2$\times$10$^{-7}$
respectively. Almost all the dwarf galaxies here obey the same
correlation as the normal galaxies do, except Mrk1434 which has the
lowest metallicity in our sample. Although the MIR ratio 
$\nu L_{\nu}[8 \mu m(dust)]/\nu L_{\nu}[24 \mu m]$ also correlates with
metallicity,  this correlation only exists for galaxies with
metallicities lower than 8.7. For galaxies with higher metallicities,
the $\nu L_{\nu}[8 \mu m(dust)]/\nu L_{\nu}[24 \mu m]$ ratios almost
remain constant. Compared with the above two correlations, the
Spearman correlation between $\nu L_{\nu}[8 \mu m(dust)]/\nu L_{\nu}[24
\mu m]$ and metallicity is much weaker, with the probability of
null-hypothesis being 0.02.

\subsection{MIR Ratios Versus AGN Distance $d_{AGN}$}

Figure~\ref{fig4} shows the MIR ratios as functions of AGN activity.
Here we used the distance $d_{AGN}$ as defined in Section 3 to
characterize AGN activity. The MIR ratio $\nu L_{\nu}[8\mu
m(dust)]/\nu L_{\nu}[3.6\mu m]$ seems to mildly increase with the
distance $d_{AGN}$. The probability of null-hypothesis of the
Spearman correlation is 0.3, indicating that the correlation is
poor. The ratio $\nu L_{\nu}[24\mu m]/\nu L_{\nu}[3.6 \mu m]$ is
well correlated with $d_{AGN}$. The probability of null-hypothesis
of such a correlation is 3.6$\times$10$^{-4}$, indicating that the
24$\mu$m emission in mass unit tend to be stronger with the level of 
AGN activity. On the contrary, the MIR ratio $\nu L_{\nu}[8 \mu
m(dust)]/\nu L_{\nu}[24 \mu m]$ is anti-correlated with $d_{AGN}$.
The probability of null-hypothesis is 1.7$\times$10$^{-5}$. The
correlation between $\nu L_{\nu}[8 \mu m(dust)]/\nu L_{\nu}[24 \mu
m]$ and $d_{AGN}$ is the tightest among the three. Even the most
discrepant point in Figure~\ref{fig4} (a) and (b) also follows such a
correlation.

\subsection{MIR Ratios Versus Internal Reddening}

To explore possible influence of the internal reddening on the MIR
emission, the MIR ratios were also plotted against E(B-V) 
in Figure~\ref{fig5}. For HII galaxies, both ratios, 
$\nu L_{\nu}[8\mu m(dust)]/\nu L_{\nu}[3.6\mu m]$ and $\nu L_{\nu}[24\mu m]/\nu L_{\nu}[3.6 \mu m]$, 
show tight correlations with E(B-V). 
The probabilities of null-hypothesis of the Spearman
correlation are 3.1$\times$10$^{-22}$ and 3.4$\times$10$^{-10}$
respectively. Both correlations indicate that the dusty galaxies
have relatively strong 8$\mu$m or 24$\mu$m dust emission. Such
correlations can also be expected in Figure~\ref{fig3}, since the
metallicities of HII galaxies are also correlated with their
internal reddening (Figure~\ref{fig2}). However, $\nu L_{\nu}[8 \mu
m(dust)]/\nu L_{\nu}[24 \mu m]$ does not show any correlation with
E(B-V). The probability of null-hypothesis is 0.16. Therefore, this
ratio does not depend on the internal reddening.

For AGNs, all the three MIR ratios do not correlate with reddening.
The Spearman correlations show that the probabilities of
null-hypothesis are 1.00, 0.16, and 0.09 respectively, 
indicating that the internal reddening does not affect all the three
MIR ratios of AGNs.

\section{DISCUSSIONS}
\label{sec discuss}

\subsection{Aperture Effect}

Since the galaxies in our sample cover a redshift range from 0.004
to 0.17, the 6\arcsec \ aperture corresponds to a physical size from
several hundred pc to several tens kpc. Will the aperture affect the
correlations we obtained? We examined this by grouping the galaxies
into several redshift bins, as shown in Figure~\ref{fig3} and
Figure~\ref{fig4}. We find that either the HII galaxies
(Figure~\ref{fig3}) or AGNs (Figure~\ref{fig4}) with different
redshifts follow the same correlations.

To further explore the aperture affect, we plotted the MIR ratios as
functions of redshift in Figure~\ref{fig6}. The mean MIR ratios (except for dwarf galaxies) 
and the corresponding standard deviations with a redshift bin of 0.02
were also plotted. There is only a small variation of the mean MIR
ratios (less than 0.2 dex) with redshift.

Meanwhile, we ''virtually'' placed a template galaxy (e.g., the SINGS
galaxy NGC~3351) in different redshifts. The 6\arcsec \ aperture
corresponds to different physical sizes at different redshifts.
Therefore, we did the aperture photometry for the MIR images of the SINGS
galaxy NGC~3351 with a set of apertures, whose physical sizes are
equal to those of the 6\arcsec \ aperture at different redshifts. As
a result, we can obtain the MIR ratios of NGC~3351 as functions of
redshift or aperture, which are plotted as the dotted curves in
Figure~\ref{fig6}. Here we only gave the MIR ratios of NGC~3351 at
the rest-frame and do not consider the K-correction. From the
figure, all the three MIR ratios of the template NGC~3351 only vary
by about 0.1 to 0.2 dex with redshift.

Both the sample statistics and template modelling show that the
aperture effect on the MIR ratios is weaker than the correlations shown in
Figure~\ref{fig3}. Therefore, we believe that the aperture effect
would not significantly affect our correlations obtained from
Figure~\ref{fig3} and Figure~\ref{fig4}.

\subsection{Metallicity and Dust Emission}

Although the VSG continuum also contributes to the $Spitzer$ 8$\mu$m
band \citep{smith07}, in HII galaxies, the strongest 7.7$\mu$m PAH
features still dominate the emission in this band. However, the VSG
continuum dominates the $Spitzer$ MIPS 24$\mu$m band. Therefore, the
correlations between the MIR ratios and metallicity in
Figure~\ref{fig3} reveal the relationships between the ratios of
PAH-to-star/VSG-to-star/PAH-to-VSG and metallicity. Here, the
3.6$\mu$m luminosity can approximately represent  the stellar mass
\citep{smithbj07,hancock07}. All these ratios increase with
metallicity, though the details are different (e.g., different
slopes). Apparently the dust properties in the HII galaxies seem to
depend on metallicity. However, the metallicity itself is often
related to many other factors, such as mass (the mass-metallicity
relation, i.e., lower mass dwarf galaxies often have lower
metallicities), and radiation field (in a low metallicity environment,
the radiation field is often hard). Hence, it is necessary to
explore the underlying mechanisms.

In fact, all these MIR behaviors mainly depend on two main factors:
the local radiation field and the dust. The properties of the
radiation field include the hardness, which can be characterized by
the MIR [NeIII]/[NeII] \citep{Thornley00,wuyl06,madden06}, and the
intensity, which can be characterized by either the MIR luminosity
density \citep{engelbracht05,wuyl06} or star formation rate
\citep{rosenberg06}. The very hard and intense radiation field could
destroy the PAHs \citep{galliano05}, and even the VSGs
\citep{contursi00}. 

The dust properties include the amount of dust,
and the fractions and spatial distributions of the PAHs and VSGs.
Galaxies with lower or vanishing PAH or VSG emission could
contain less amount of dust, possibly because they have low
metallicities, which lack material to form dust, or their masses are too low
to retain the dust against the radiation pressure and winds, 
or they are too young, to have had enough time to form PAHs \citep{hogg05,madden06,draine07b}. 
Different fractions and distributions of the PAHs and VSGs could result 
in different PAH-to-VSG ratios.

\begin{figure}
\centerline{\includegraphics[height=0.9\textwidth]{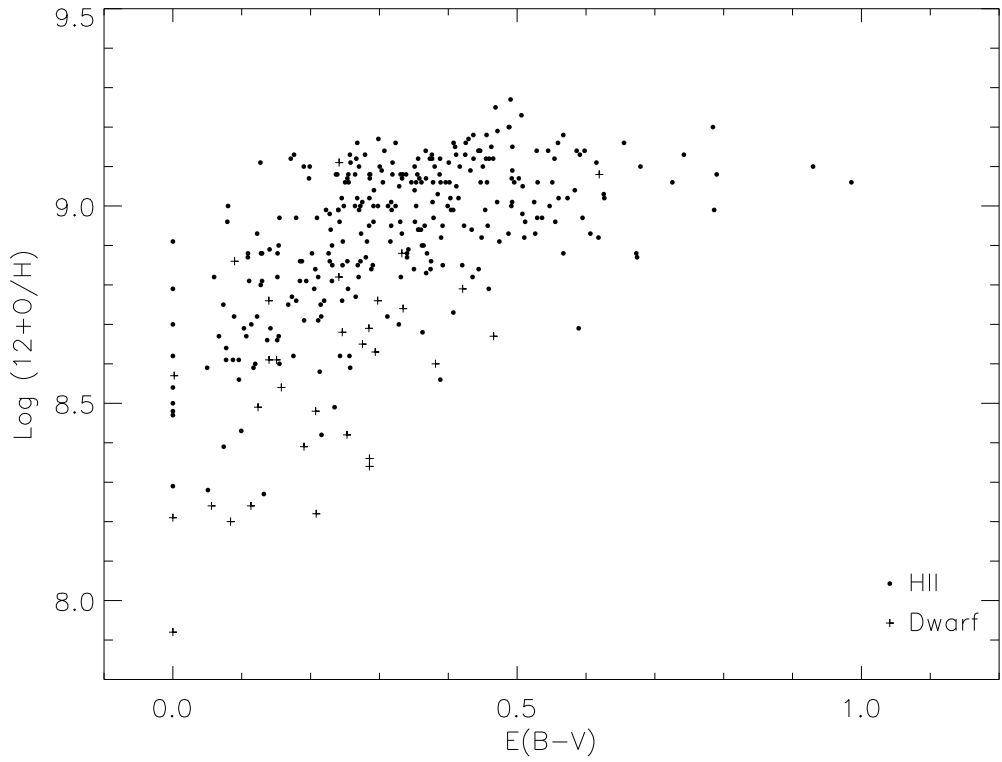}}
\caption{The metallicities of HII galaxies as a function of their internal reddening.
The plus symbols are dwarf galaxies.
} \label{fig2}
\end{figure}

\begin{figure}
\centerline{\includegraphics[height=0.9\textwidth]{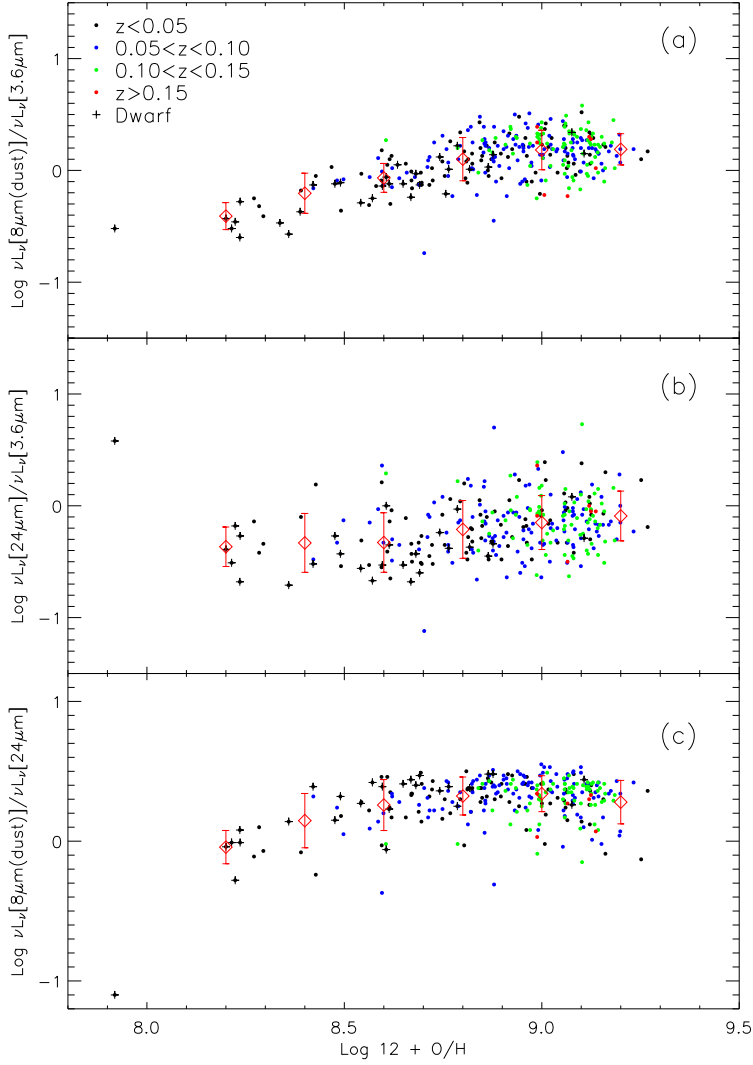}}
\caption{The MIR ratios (a) $\nu L_{\nu}[8\mu m(dust)]/\nu L_{\nu}[3.6\mu m]$, ]
(b) $\nu L_{\nu}[24\mu m]/\nu L_{\nu}[3.6 \mu m]$,
and (c) $\nu L_{\nu}[8 \mu m(dust)]/\nu L_{\nu}[24 \mu m]$ versus
metallicity. The plus symbols represent the dwarf galaxies and the
solid circles are normal galaxies. The diamonds and bars give the mean
values and 1-$\sigma$ standard deviations with a metallicity bin of
0.2. Different colors show galaxies in different redshift bins.} 
\label{fig3}
\end{figure}

\begin{figure}
\centerline{\includegraphics[height=0.9\textwidth]{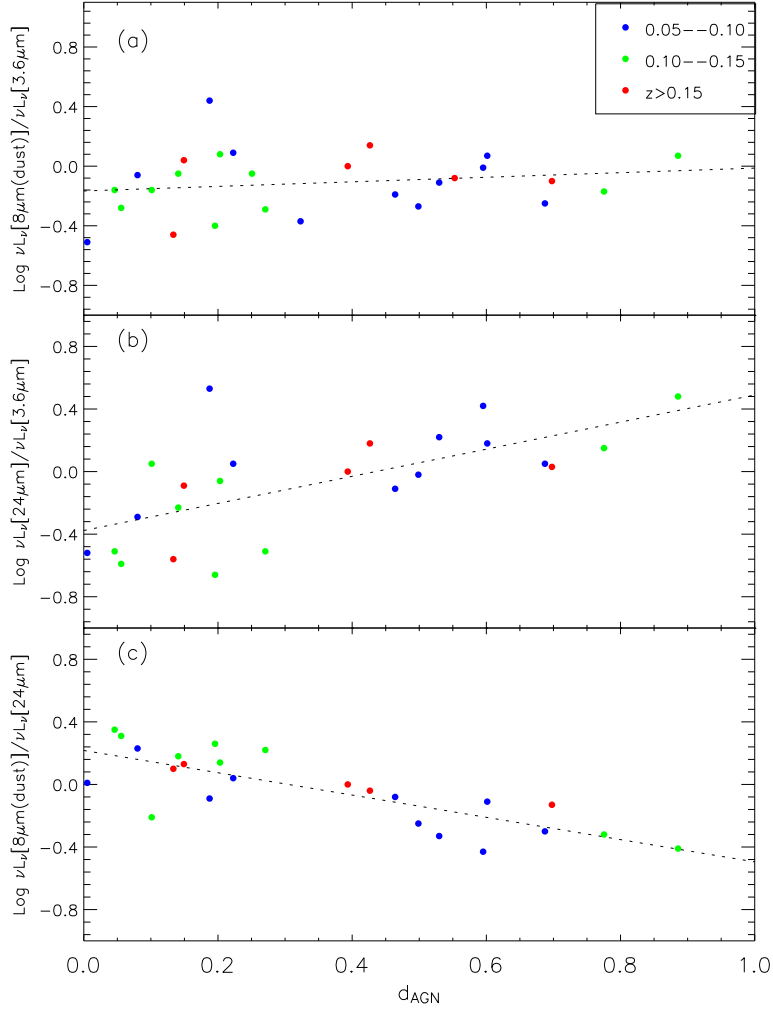}}
\caption{The MIR  ratios (a) $\nu L_{\nu}[8\mu m(dust)]/\nu L_{\nu}[3.6\mu m]$, 
(b) $\nu L_{\nu}[24\mu m]/\nu L_{\nu}[3.6 \mu m]$,
and (c) $\nu L_{\nu}[8 \mu m(dust)]/\nu L_{\nu}[24 \mu m]$ versus
the distance $d_{AGN}$. Different colors represent AGNs in different
redshift bins. The dotted lines represent the best linear fitting. }
\label{fig4}
\end{figure}

\begin{figure}
\centerline{\includegraphics[height=0.9\textwidth]{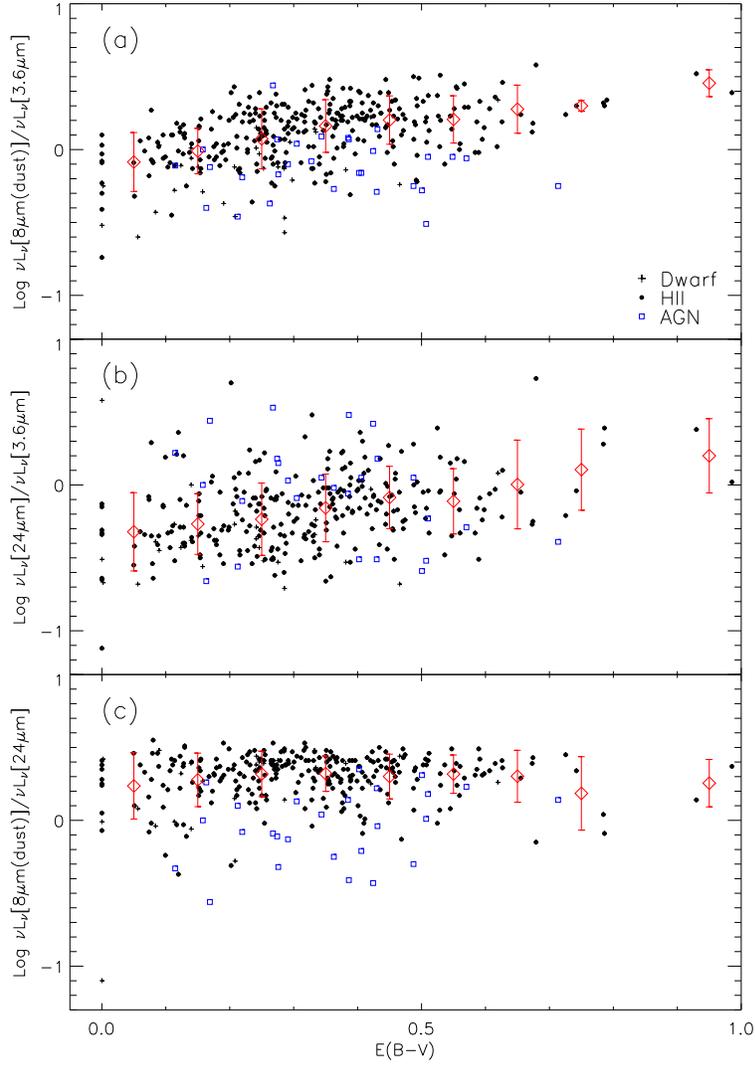}}
\caption{The MIR  ratios (a) $\nu L_{\nu}[8\mu m(dust)]/\nu L_{\nu}[3.6\mu m]$,
(b) $\nu L_{\nu}[24\mu m]/\nu L_{\nu}[3.6 \mu m]$,
and (c) $\nu L_{\nu}[8 \mu m(dust)]/\nu L_{\nu}[24 \mu m]$
as functions of reddening. The symbols are the same as in Figure~\ref{fig1}.
The diamonds and bars give the mean values and 1-$\sigma$ standard deviations 
of HII galaxies with an E(B-V) bin of 0.1.
} \label{fig5}
\end{figure}

\begin{figure}
\centerline{\includegraphics[height=0.9\textwidth]{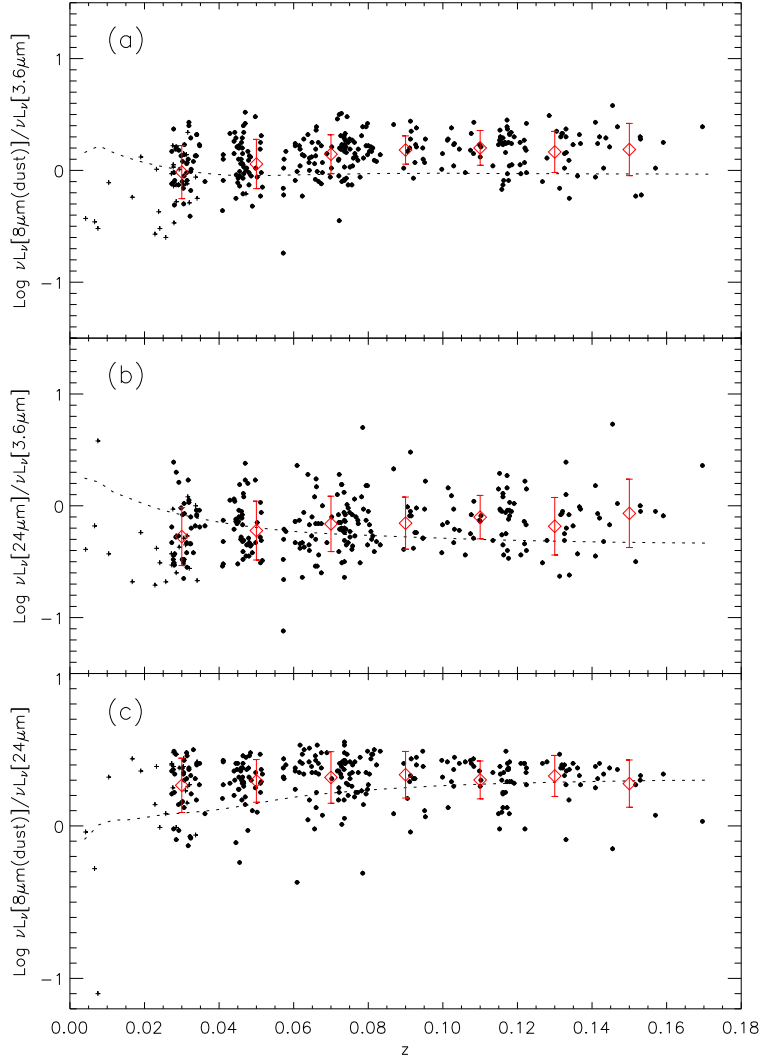}}
\caption{The MIR  ratios (a) $\nu L_{\nu}[8\mu m(dust)]/\nu L_{\nu}[3.6\mu m]$,
(b) $\nu L_{\nu}[24\mu m]/\nu L_{\nu}[3.6 \mu m]$,
and (c) $\nu L_{\nu}[8 \mu m(dust)]/\nu L_{\nu}[24 \mu m]$
as functions of redshift.
The diamonds and bars are the mean values and 1-$\sigma$ standard deviations with a redshift bin
of 0.02.
The dotted curves show the variation of the MIR ratios of NGC~3351 with
redshift (aperture). Dward galaxies are labelled as plus symbols.
} \label{fig6}
\end{figure}

The positive correlations between the PAH/VSG-to-star ratios and
metallicity obtained from Figure~\ref{fig3} could be explained by the
amount of dust. It could be expected from the relations between the
reddening and metallicity (Figure~\ref{fig2}) or the PAH/VSG-to-star
ratios (Figure~\ref{fig5} (a) and (b)). In fact, the internal reddening
is directly related to the amount of dust if assuming a
constant dust-to-gas ratio, due to the tight correlation between E(B-V)
and H column density \citep{bohlin78,draine03}. Therefore, the low
PAH/VSG-to-star ratio of lower metallicity galaxies could not be due to
that their masses are too low to retain the amount of dust, because
some of the normal galaxies also follow the same correlation 
as lower metallicity galaxies do in Figure~\ref{fig3}. 
It could be that they are so young that a large
amount of dust has not had enough time to be released. This is
supported by Figure~\ref{fig2}. 

As for the PAH-to-VSG ratio, if
assuming that the AGB stars are the dominant carbon reservoir to
form the PAHs \citep{dwek98} with an age of star formation less than
500Myr when the 4M$ {\odot}$ has entered the AGB stage\citep{madden06}, 
the PAH-to-VSG ratio could increase with time.  After that, the ratio
could approach a constant. This seems to be consistent with
Figure~\ref{fig3}(c), in which the value of $\nu L_{\nu}[8 \mu
m(dust)]/\nu L_{\nu}[24 \mu m]$ increases with metallicity and then
remains constant beyond solar metallicity. Such an explanation is
quite similar to the explanation of the low dust-to-gas ratios for
low metallicity SINGS galaxies by \citet{draine07b}. 
Also the behavior of the PAH-to-VSG ratio with time can be well described 
by the Figure 4 of \citet{galliano06} recent work on 
chemical evolution modeling of carbon and silicate grains.

\begin{figure}
\centerline{\includegraphics[height=0.9\textwidth]{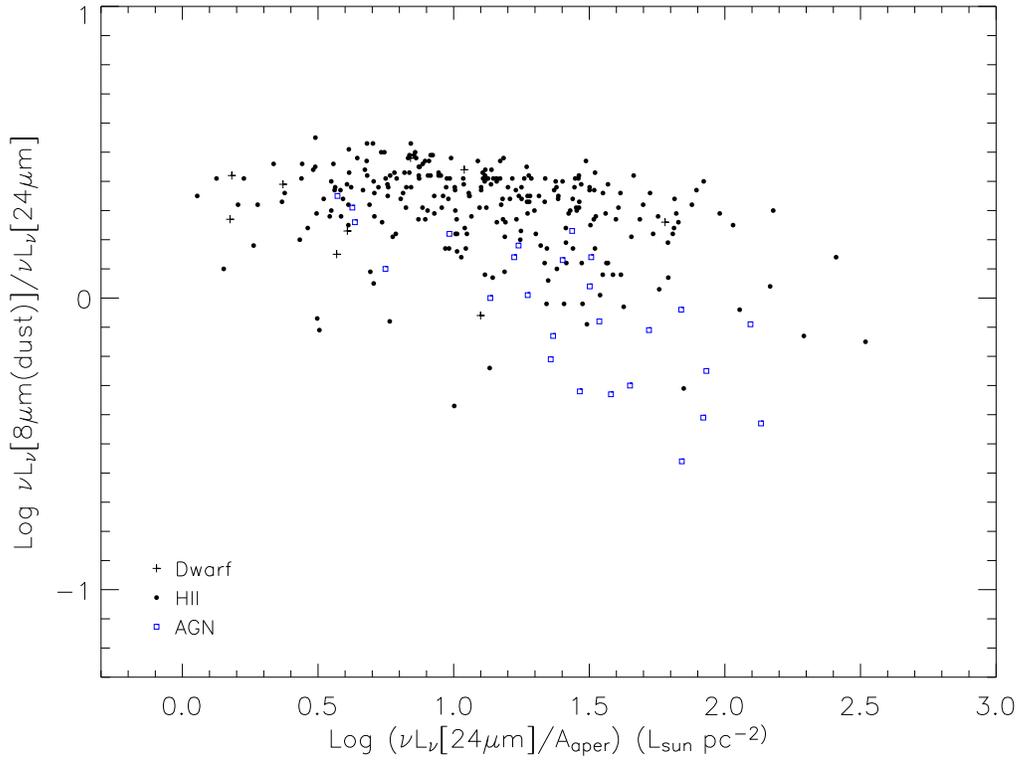}}
\caption{The MIR  ratio $\nu L {\nu}[8 \mu m(dust)]/\nu L {\nu}[24 \mu m]$
as a function of the 24$\mu$m intensity. The 24$\mu$m intensity is defined as
the 24$\mu$m luminosity in unit physical area and is obtained from the ratio of
the 24$\mu$m luminosity to the physical area of aperture 6 \arcsec \ in
each galaxy.
The symbols are the same as in Figure~\ref{fig1}.
} \label{fig7}
\end{figure}

Another explanation could be that the hard and intense radiation
field from the young stars in the lower metallicity galaxies can
destroy not only the PAHs but also the VSGs. Since the PAHs are much
more sensitive to hard radiation field than the VSGs are, the slope of
the PAH-metallicity relation is much steeper than that of the VSGs.
This results in a decreasing PAH-to-VSG ratio in the lower
metallicity galaxies (Figure~\ref{fig3}). Such a destruction effect
is further supported by Figure~\ref{fig7}, which shows the
anti-correlation between $\nu L_{\nu}[8 \mu m(dust)]/\nu L_{\nu}[24
\mu m]$ and the 24$\mu$m intensity for the sample of HII galaxies,
though such an anti-correlation is not as steep as that for AGNs.
The 24$\mu$m intensity is obtained from the ratio of the 24$\mu$m
luminosity to the physical area in the aperture of 6\arcsec. 

We thus conclude that both effects of dust formation and destruction can
explain the correlations between the MIR ratios and metallicity.
This is consistent with the conclusions for PAHs by \citet{wuyl06}
and \citet{draine07a}.

\subsection{AGN Activity and Dust Emission}

The MIR spectra of AGNs were already noted to be void of PAH
features \citep{roche85, roche91, aitken85, genzel00, laurent00,
siebenmorgen04}. This could be explained by the destruction of PAHs
due to the hard radiation field of the central source
\citep{madden06}.  It is also supported by recent $Spitzer$ results
given by \citet{dale06} and \citet{armus07}, who used [OIV]/[NeII]
as the indicator of the hardness of radiation field.  However,
Figure~\ref{fig4}(a) shows that the PAH-to-star ratio apparently
does not have any tendency to decrease but rather increases with AGN
activity. \citet{draine07b} explained the SINGS AGNs as
low-luminosity AGNs, which have little effect on the 8$\mu$m dust
emission so that they show no evidence of PAH suppression. This is
however not the case for our sample of AGNs, since all of them have
higher H$ {\alpha}$ equivalent widths according to our selection in Section
2. The amount of dust cannot account for that, since the internal
reddening does not correlate with any of the three MIR ratios
(Figure~\ref{fig5}). We suggest that the AGNs not only can destroy
the PAHs in the central regions but could also possibly excite the
PAH emission in the outer regions \citep{smith07}.

Another explanation is that the central AGN raises the level of the
VSG continuum in the 8$\mu$m band. Similarly the sharp increase of
the VSG-to-star ratio with AGN activity in Figure~\ref{fig4}(b) may
also be due to the enhanced 24$\mu$m VSG continuum by the powerful AGN.
Although the outer PAH emission could be excited by the central AGN,
both the destruction of PAHs and the enhancement of the 24$\mu$m VSG continuum
by the AGN would result in the tight anti-correlation
between the PAH-to-VSG ratio and AGN activity.

\subsection{Radiation Field and Dust Emission}

One of the important factors in explaining the correlations between
the MIR ratios and metallicity or AGN activity is the radiation
field, which could be provided by either young stars or AGNs.
\citet{hogg05} suggested that [OIII]/H$ {\beta}$  could be an
indicator of radiation field hardness. 

\begin{figure}
\centerline{\includegraphics[height=0.9\textwidth]{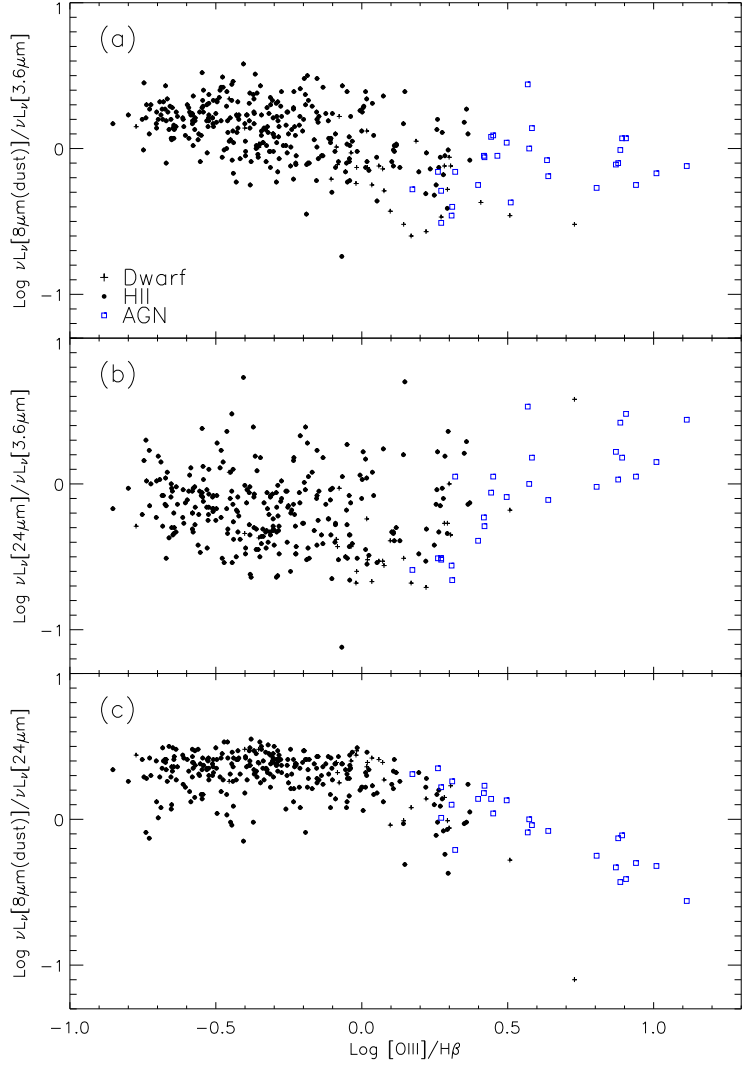}}
\caption{The MIR  ratios (a) $\nu L_{\nu}[8\mu m(dust)]/\nu L_{\nu}[3.6\mu m]$,
(b) $\nu L_{\nu}[24\mu m]/\nu L_{\nu}[3.6 \mu m]$,
and (c) $\nu L_{\nu}[8 \mu m(dust)]/\nu L_{\nu}[24 \mu m]$
as functions of the emission line ratio log([OIII]/H$ {\beta}$).
The symbols are the same as in Figure~\ref{fig1}.
} \label{fig8}
\end{figure}

Figure~\ref{fig8} shows the
three MIR ratios as functions of [OIII]/H$ {\beta}$. For AGNs, the
correlations between the MIR ratios and [OIII]/H$ {\beta}$ are
similar to those between the MIR ratios and $d_{AGN}$. So, the AGN
distance $d_{AGN}$ is directly related to the radiation field. For
HII galaxies, all these MIR ratios show the anti-correlation with
[OIII]/H$ {\beta}$. With [OIII]/H$ {\beta}$ increasing, the
PAH/VSG-to-star ratios of the HII galaxies and AGNs show different
features. Hence, the PAH/VSG-to-star ratios seem to depend on not
only the radiation field, but also some other important factor as
well. As discussed above, in the HII galaxies. the amount of dust 
would play an important role. 
However the PAH-to-VSG ratio, which 
describes the fraction of two different dust components, does not 
depend on the amount of dust but rather the radiation field. So, the
PAH-to-VSG ratio of the HII galaxies and AGNs would present the similar behaviors
along [OIII]/H$ {\beta}$, as shown in Figure~\ref{fig8} (c).

In a hard radiation field environment ([OIII]/H$ {\beta}$ $>$ 1),
the PAH-to-VSG ratio decreases sharply with increasing [OIII]/H$
{\beta}$, whether in HII galaxies or in AGNs. However, the PAH-to-VSG
ratio remains constant in a weak radiation field environment
([OIII]/H$ {\beta}$ $<$ 1). We conclude that the hard radiation field, 
rather than metallicity or AGN activity, is the most direct factor to determine
the PAH-to-VSG ratio.

\section{SUMMARY}

We have constructed a sample of galaxies from both the SDSS Data
Released 4 and the $Spitzer$  SWIRE fields to study possible effects
of both metallicity and AGN activity on the MIR dust emission. 
Based on this sample we have found that:

The MIR ratios: PAH-to-star, VSG-to-star and PAH-to-VSG, which can
be characterized respectively by $\nu L_{\nu}[8\mu m(dust)]/\nu L_{\nu}[3.6\mu m]$, 
$\nu L_{\nu}[24\mu m]/\nu L_{\nu}[3.6\mu m]$, \\  and
$\nu L_{\nu}[8\mu m(dust)]/\nu L_{\nu}[24\mu m]$, are found to be
positively correlated with the metallicities of galaxies. The above
correlations could be explained by either the amount of dust (ongoing
dust formation) or dust destruction (either the PAHs or VSGs could
be destroyed by the hard and strong radiation field).

The VSG-to-star and PAH-to-VSG ratios are strongly correlated or
anti-correlated with AGN activity.  However, the PAH-to-star ratio
poorly depends on AGN activity. This may be due to the PAH destruction
or enhanced VSG continuum by the central AGN.

\section{ACKNOWLEDGEMENTS}

We thank Jia-sheng Huang, Zhong Wang, David Sanders, and Yan-Chun Liang
for helpful discussions. Also many thanks to anonymous referees. 
H.W. is grateful to the Institute for Astronomy, University of Hawaii for hospitality. 
This work was
supported by the National Science Foundation of China under grants
10273012, 10333060, and 10473013, and a CAS grant KJCX3-SYW-N2. 
It is, in part, based on observations made with the Spitzer Space Telescope, which is
operated by Jet Propulsion Laboratory of the California Institute of
Technology under NASA Contract 1407.

Funding for the SDSS and SDSS-II has been provided by the Alfred
P. Sloan Foundation, the Participating Institutions, the National Science
Foundation, the U.S. Department of Energy, the National Aeronautics
and Space Administration, the Japanese Monbukagakusho, the Max Planck
Society, and the Higher Education Funding Council for England. The SDSS Web Site is http://www.sdss.org/.

The SDSS is managed by the Astrophysical Research Consortium for the Participating Institutions. The Participating Institutions are the American Museum of Natural History, Astrophysical Institute Potsdam, University of Basel, University of Cambridge, Case Western Reserve University, University of Chicago, Drexel University, Fermilab, the Institute for Advanced Study, the Japan Participation Group, Johns Hopkins University, the Joint Institute for Nuclear Astrophysics, the Kavli Institute for Particle Astrophysics and Cosmology, the Korean Scientist Group, the Chinese Academy of Sciences (LAMOST), Los Alamos National Laboratory, the Max-Planck-Institute for Astronomy (MPIA), the Max-Planck-Institute for Astrophysics (MPA), New Mexico State University, Ohio State University, University of Pittsburgh, University of Portsmouth, Princeton University, the United States Naval Observatory, and the University of Washington.

\clearpage

\clearpage

\clearpage

\clearpage

\clearpage

\clearpage

\clearpage

\clearpage

\begin{thebibliography}{99}

\bibitem[Adelman-McCarthy et al.(2006)]{adelman06} Adelman-McCarthy, J.~K., et al.\ 2006, \apjs, 162, 38
\bibitem[Aitken \& Roche(1985)]{aitken85} Aitken, D.~K., \& Roche, P.~F.\ 1985, \mnras, 213, 777
\bibitem[Alonso-Herrero et al.(2006)]{alonso06} Alonso-Herrero, A., Rieke, G.~H., Rieke, M.~J., Colina, L., P{\'e}rez-Gonz{\'a}lez, P.~G., \& Ryder, S.~D.\ 2006, \apj, 650, 835
\bibitem[Armus et al.(2007)]{armus07} Armus, L., et al.\ 2007, \apj, 656, 148
\bibitem[Bertin \& Arnouts(1996)]{bertin96} Bertin, E., \& Arnouts, S.\ 1996, \aaps, 117, 393
\bibitem[Bohlin et al.(1978)]{bohlin78} Bohlin, R.~C., Savage, B.~D., \& Drake, J.~F.\ 1978, \apj, 224, 132
\bibitem[Boselli et al.(2004)]{boselli04} Boselli, A., Lequeux, J., \& Gavazzi, G.\ 2004, \aap, 428, 409
\bibitem[Bruzual \& Charlot(2003)]{bc03} Bruzual, G., \& Charlot, S.\ 2003, \mnras, 344, 1000
\bibitem[Calzetti et al.(2005)]{calzetti05} Calzetti, D., et al.\ 2005, \apj, 633, 871
\bibitem[Calzetti(2001)]{calzetti01} Calzetti, D.\ 2001, \pasp, 113, 1449
\bibitem[Cao \& Wu(2007)]{cao07} Cao, C., \& Wu, H.\ 2007, \aj, 133, 1710
\bibitem[Contursi et al.(2000)]{contursi00} Contursi, A., et al.\ 2000, \aap, 362, 310
\bibitem[Cutri et al.(2003)]{cutri03} Cutri, R.~M., et al.\ 2003, The IRSA 2MASS All-Sky Point Source Catalog, NASA/IPAC
\bibitem[Dale et al.(2006)]{dale06} Dale, D.~A., et al.\ 2006, \apj, 646, 161
\bibitem[Draine \& Li(2007)]{draine07a} Draine, B.~T., \& Li, A.\ 2007, \apj, 657, 810
\bibitem[Draine et al.(2007)]{draine07b} Draine, B.~T., et al.\ 2007, ArXiv Astrophysics e-prints, arXiv:astro-ph/0703213
\bibitem[Draine(2003)]{draine03} Draine, B.~T.\ 2003, \araa, 41, 241
\bibitem[Dwek(1998)]{dwek98} Dwek, E.\ 1998, \apj, 501, 643
\bibitem[Elbaz et al.(2002)]{elbaz02} Elbaz, D., Cesarsky, C.~J., Chanial, P., Aussel, H., Franceschini, A., Fadda, D., \& Chary, R.~R.\ 2002, \aap, 384, 848
\bibitem[Engelbracht et al.(2005)]{engelbracht05} Engelbracht, C.~W., Gordon, K.~D., Rieke, G.~H., Werner, M.~W., Dale, D.~A., \& Latter, W.~B.\ 2005, \apjl, 628, L29
\bibitem[Fazio et al.(2004)]{fazio04} Fazio, G. G. et al., 2004, \apjs, 154, 10
\bibitem[F{\"o}rster Schreiber et al.(2004)]{forster04} F{\"o}rster Schreiber, N.~M., Roussel, H., Sauvage, M., \& Charmandaris, V.\ 2004, \aap, 419, 501
\bibitem[Galliano et al.(2005)]{galliano05} Galliano, F., Madden, S.~C., Jones, A.~P., Wilson, C.~D., \& Bernard, J.-P.\ 2005, \aap, 434, 867
\bibitem[Galliano(2006)]{galliano06} Galliano, F.\ 2006, ArXiv Astrophysics e-prints, arXiv:astro-ph/0610852 
\bibitem[Genzel \& Cesarsky(2000)]{genzel00} Genzel, R., \& Cesarsky, C.~J.\ 2000, \araa, 38, 761
\bibitem[Hancock et al.(2007)]{hancock07} Hancock, M., Smith, B.~J., Struck, C., Giroux, M.~L., Appleton, P.~N., Charmandaris, V., \& Reach, W.~T.\ 2007, \aj, 133, 676
\bibitem[Hogg et al.(2005)]{hogg05} Hogg, D.~W., Tremonti, C.~A., Blanton, M.~R., Finkbeiner, D.~P., Padmanabhan, N., Quintero, A.~D., Schlegel, D.~J., \& Wherry, N.\ 2005, \apj, 624, 162
\bibitem[Houck et al.(2004)]{houck04a} Houck, J.~R., et al.\ 2004a, \apjs, 154, 18 
\bibitem[Houck et al.(2004)]{houck04b} Houck, J.~R., et al.\ 2004b, \apjs, 154, 211
\bibitem[Huang et al.(2004)]{huang04} Huang, J.-S., et al.\ 2004, \apjs, 154, 44
\bibitem[Kauffmann et al.(2003)]{kauffmann03} Kauffmann, G., et al.\ 2003, \mnras, 346, 1055
\bibitem[Kennicutt et al.(2003)]{kennicutt03} Kennicutt, R.~C., Jr., et al.\ 2003, \pasp, 115, 928
\bibitem[Kewley et al.(2001)]{kewley01} Kewley, L.~J., Dopita, M.~A., Sutherland, R.~S., Heisler, C.~A., \& Trevena, J.\ 2001, \apj, 556, 121
\bibitem[Kewley et al.(2006)]{kewley06} Kewley, L.~J., Groves, B., Kauffmann, G., \& Heckman, T.\ 2006, \mnras, 372, 961
\bibitem[Lacy et al.(2005)]{lacy05} Lacy, M., et al.\ 2005, \apjs, 161, 41
\bibitem[Laurent et al.(2000)]{laurent00} Laurent, O., Mirabel, I.~F., Charmandaris, V., Gallais, P., Madden, S.~C., Sauvage, M., Vigroux, L., \& Cesarsky, C.\ 2000, \aap, 359, 887
\bibitem[Lonsdale et al.(2003)]{lonsdale03} Lonsdale, C.~J., et al.\ 2003, \pasp, 115, 897
\bibitem[Madden(2000)]{madden00} Madden, S.~C.\ 2000, New Astronomy Review, 44, 249
\bibitem[Madden et al.(2006)]{madden06} Madden, S.~C., Galliano, F., Jones, A.~P., \& Sauvage, M.\ 2006, \aap, 446, 877
\bibitem[Oke \& Gunn(1983)]{oke83} Oke, J.~B., \& Gunn, J.~E.\ 1983, \apj, 266, 713
\bibitem[Pagel \& Edmunds(1981)]{pagel81} Pagel, B.~E.~J., \& Edmunds, M.~G.\ 1981, \araa, 19, 77
\bibitem[Rieke et al.(2004)]{rieke04} Rieke, G.~H., et al.\ 2004, \apjs, 154, 25
\bibitem[Roche et al.(1991)]{roche91} Roche, P.~F., Aitken, D.~K., Smith, C.~H., \& Ward, M.~J.\ 1991, \mnras, 248, 606
\bibitem[Roche \& Aitken(1985)]{roche85} Roche, P.~F., \& Aitken, D.~K.\ 1985, \mnras, 213, 789
\bibitem[Rosenberg et al.(2006)]{rosenberg06} Rosenberg, J.~L., Ashby, M.~L.~N., Salzer, J.~J., \& Huang, J.-S.\ 2006, \apj, 636, 742
\bibitem[Roussel et al.(2001)]{roussel01} Roussel, H., Sauvage, M., Vigroux, L., \& Bosma, A.\ 2001, \aap, 372, 427
\bibitem[Siebenmorgen et al.(2004)]{siebenmorgen04} Siebenmorgen, R., Kr{\"u}gel, E., \& Spoon, H.~W.~W.\ 2004, \aap, 414, 123
\bibitem[Smith et al.(2007a)]{smithbj07} Smith, B.~J., Struck, C., Hancock, M., Appleton, P.~N., Charmandaris, V., \& Reach, W.~T.\ 2007a, \aj, 133, 791
\bibitem[Smith et al.(2007b)]{smith07} Smith, J.~D.~T., et al.\ 2007b, \apj, 656, 756
\bibitem[Smith et al.(2002)]{smith02} Smith, J.~A., et al.\ 2002, \aj, 123, 2121
\bibitem[Stoughton et al.(2002)]{stoughton02} Stoughton, C., et al.\ 2002, \aj, 123, 485
\bibitem[Sturm et al.(2002)]{sturm02} Sturm, E., Lutz, D., Verma, A., Netzer, H., Sternberg, A., Moorwood, A.~F.~M., Oliva, E., \& Genzel, R.\ 2002, \aap, 393, 821
\bibitem[Surace et al.(2005)]{surace05} Surace J.~A., et al., 2005, The SWIRE Data Release 2: Image Atlases and Source Catalogs for ELAIS-N1, ELAIS-N2, XMM-LSS, and the Lockman Hole
\bibitem[Thornley et al.(2000)]{Thornley00} Thornley, M.~D., Schreiber, N.~M.~F., Lutz, D., Genzel, R., Spoon, H.~W.~W., Kunze, D., \& Sternberg, A.\ 2000, \apj, 539, 641
\bibitem[Thuan et al.(1999)]{thuan1999} Thuan, T.~X., Sauvage, M., \& Madden, S.\ 1999, \apj, 516, 783
\bibitem[Tremonti et al.(2004)]{tremonti04} Tremonti, C.~A., et al.\ 2004, \apj, 613, 898
\bibitem[Veilleux \& Osterbrock(1987)]{veilleux87} Veilleux, S., \& Osterbrock, D.~E.\ 1987, \apjs, 63, 295
\bibitem[Verma et al.(2005)]{verma2005} Verma, A., Charmandaris, V., Klaas, U., Lutz, D., \& Haas, M.\ 2005, Space Science Reviews, 119, 355
\bibitem[Vogler et al.(2005)]{vogler05} Vogler, A., Madden, S.~C., Beck, R., Lundgren, A.~A., Sauvage, M., Vigroux, L., \& Ehle, M.\ 2005, \aap, 441, 491
\bibitem[Weedman et al.(2005)]{weedman05} Weedman, D.~W., et al.\ 2005, \apj, 633, 706
\bibitem[Wen et al.(2007)]{wen07} Wen, X.~Q., Wu, H., Cao, C., Xia, X.~Y., 2007, ChJAA, 7, 187
\bibitem[Werner et al.(2004)]{werner04} Werner, M. W. et al., 2004, \apjs, 154, 1
\bibitem[Wu et al.(1998)]{wu98} Wu, H., Zou, Z.~L., Xia, X.~Y., \& Deng, Z.~G.\ 1998, \aaps, 132, 181
\bibitem[Wu et al.(2005)]{wu05} Wu, H., Cao, C., Hao, C.-N., Liu, F.-S., Wang, J.-L., Xia, X.-Y., Deng, Z.-G., \& Young, C.~K.-S.\ 2005, \apjl, 632, L79
\bibitem[Wu et al.(2006)]{wuyl06} Wu, Y., Charmandaris, V., Hao, L., Brandl, B.~R., Bernard-Salas, J., Spoon, H.~W.~W., \& Houck, J.~R.\ 2006, \apj, 639, 157

\end{thebibliography}
\end{document}